\documentclass[twocolumn,english,superscriptaddress,amsmath,amssymb,floats,prb,longbibliography]{revtex4-2}
\usepackage[latin9]{inputenc}
\usepackage{color}
\usepackage{amsmath}
\usepackage{amssymb}
\usepackage{graphicx}
\graphicspath{ {./figures5} }
\usepackage{esint}
\usepackage[normalem]{ulem}

\makeatletter
\usepackage{color}
\usepackage{wasysym}
\usepackage{physics}
\renewcommand\[{\begin{equation}}
\renewcommand\]{\end{equation}}
\usepackage[unicode=true,
 bookmarks=false,
 breaklinks=false,pdfborder={0 0 1}]
 {hyperref}

\newcommand{\mbf}[1]{\mathbf{#1}}

\usepackage{hyperref}
\hypersetup{
	colorlinks=false,
	linkcolor=black,      
	urlcolor=black,
	citecolor=black,
}

\makeatother

\begin{document}
\title{Hot-lines topology and the fate of the spin resonance mode in three-dimensional unconventional superconductors}
\author{Fei Chen}
\affiliation{School of Physics and Astronomy, University of Minnesota, Minneapolis,
55455, USA}
\author{Rafael M. Fernandes}
\affiliation{School of Physics and Astronomy, University of Minnesota, Minneapolis,
55455, USA}
\author{Morten H. Christensen}
\affiliation{Niels Bohr Institute, University of Copenhagen, 2100 Copenhagen, Denmark}
\date{\today}

\begin{abstract}
In the quasi-two-dimensional (quasi-2D) copper- and iron-based superconductors, the onset of superconductivity is accompanied by a prominent peak in the magnetic spectrum at momenta close to the wave-vector of the nearby antiferromagnetic state. Such a peak is well described in terms of a spin resonance mode, i.e., a spin-1 exciton theoretically predicted for quasi-2D superconductors with a sign-changing gap. The same theories, however, indicate that such a resonance mode should be absent in a three-dimensional (3D) system with a spherical Fermi surface. This raises the question of the fate of the spin resonance mode in layered unconventional superconductors that are not strongly anisotropic, such as certain heavy-fermion compounds and potentially the newly discovered nickelate superconductor NdNiO$_2$. Here, we use the random-phase-approximation to calculate the dynamical spin susceptibility of 3D superconductors with a $d_{x^2-y^2}$-wave gap symmetry and corrugated cylindrical-like Fermi surfaces. By varying the out-of-plane hopping anisotropy $t_z/t$, we demonstrate that the appearance of a spin resonance mode is determined by the topology of the hot lines -- i.e. lines on the Fermi surface that are connected by the magnetic wave-vector. For an in-plane antiferromagnetic wave-vector, the hot lines undergo a topological transition from open lines to closed loops at a critical $t_z/t$ value. The closed hot lines cross the nodal superconducting lines, making the spin resonance mode overdamped and incoherent. In contrast, for an out-of-plane antiferromagnetic wave-vector, the hot lines remain open and the spin resonance mode remains sharp. We discuss the experimental implications of our results for the out-of-plane dispersion of the spin resonance mode and, more generally, for inelastic neutron scattering experiments on unconventional superconductors. 
\end{abstract}
\maketitle

\section{Introduction}

The microscopic origin of the pairing interaction in unconventional superconductors is one of the outstanding questions in contemporary condensed matter physics. The phase diagrams of many unconventional superconductors feature a close proximity between magnetic and superconducting phases~\cite{norman2011challenge,scalapino2012,Keimer2015From,Fernandes2022}, hinting at the importance of magnetic fluctuations for driving the formation of a superconducting state. Indeed, phase sensitive measurements of the superconducting gap structure in the cuprates reveal a sign-change between regions of the Fermi surface connected by $\mbf{Q}$, the ordering vector of the nearby magnetic phase~\cite{Harlingen1995Phase-sensitive,Tsuei2000Pairing}, characteristic of a $d_{x^2-y^2}$-wave gap. While these types of measurements have not been carried out for the iron-based systems, indirect measurements of the phase difference between gaps of opposite bands using scanning tunnelling microscopy also indicate a sign-changing order parameter, but with an $s^{\pm}$ structure ~\cite{Hanaguri,Du2018}.

In such quasi-two-dimensional (quasi-2D) systems, a sign-changing order parameter is theoretically expected to give rise to a sharp peak in the dynamical spin susceptibility near the antiferromagnetic wave-vector $\mbf{Q}$, signalling the appearance of a so-called spin resonance mode -- i.e. a spin-1 exciton -- inside the superconducting state \cite{Norman2000,Abanov2003,Eremin2008Feedback,Eschrig2006Effect,scalapino2012}. Inelastic neutron scattering experiments provide a reliable method for assessing this peak, which is routinely observed in cuprates~\cite{Rossat-Mignod1991Neutron,Mook1993Polarized,Fong1995Phonon,Fong1999Neutron}, heavy-fermion compounds~\cite{Sato2001Strong,Stock2008Spin,Stock2008Spin,Friemel2012Resonant}, and iron-based superconductors~\cite{Christianson2008Unconventional,Inosov2010Normal-state,Dai2015Antiferromagnetic,Inosov2016Spin,Bang2017Superconducting}. While other scenarios have been proposed that also predict a peak in the magnetic spectrum inside the superconducting phase \cite{Morr1998,Demler1998Pi,Onari2010Structure,Onari2011Neutron}, some of which do not require a sign-changing gap, the peak in these cases is generally either located above twice the superconducting gap $2\Delta$ \cite{Tchernyshyov2001Neutron} or, at the very least, has an energy that is not necessarily below $2\Delta$ \cite{chubukov2008spin}. In contrast, in the spin-exciton scenario, the energy of the peak is always below $2\Delta$, and a sign-changing gap is required. For these reasons, the presence of a spin resonance mode at energies below twice the gap is often invoked as evidence for a sign-change in the order parameter~\cite{scalapino2012,Eschrig2006Effect}. 

An interesting questions that has been the subject of some debate is whether a sign-changing gap necessarily implies a spin resonance mode, particularly in systems that are not quasi-2D. Consider, for instance, the heavy-fermion compound CeCoIn$_5$, which was shown to exhibit a peak in the dynamical susceptibility below the superconducting critical temperature~\cite{Stock2008Spin,Song2020Nature}. Subsequent theoretical work argued that, due to the three-dimensional (3D) nature of CeCoIn$_5$, the resonance peak was more likely caused by the reemergence of a magnon that ceases to be overdamped once the system becomes superconducting~\cite{chubukov2008spin}. This is because a spin-resonance mode is absent in a 3D system with a spherical Fermi surface, as shown in Ref.~\onlinecite{chubukov2008spin}. The magnon-based mechanism is distinct from the spin-1 exciton scenario in that it does not necessarily require a sign-changing gap. The discovery of superconductivity in the doped infinite-layer nickel-based compound NdNiO$_2$~\cite{li2019nickelate1st,Osada2021Nickelate,Li2020Superconducting,Zeng2020Phase} brings this issue back to the discussion table. In contrast to the cuprates, the electronic structure of the nickelates features a more considerable dependence on the out-of-plane momentum $k_z$~\cite{Lecherman2020Late,Karp2020Many-body}. Moreover, no long-range magnetic order has been observed~\cite{Lu2021Magnetic}. The dichotomy between 2D and 3D systems described above motivates a further systematic study of the impact of dimensionality on the appearance and robustness of the spin resonance mode in the regime between a purely 2D and an isotropic 3D electronic structure.

In this paper, we accomplish this by computing the spin susceptibility of  a $d_{x^2y^2}$-wave superconductor with a 3D electronic structure featuring a single corrugated cylindrical-like Fermi surface. This choice is motivated by the generic dispersion of layered unconventional superconductors, and does not refer to a specific material. To contrast the quasi-2D from the less-anisotropic 3D cases in a systematic manner, we treat the ratio between the out-of-plane and in-plane hopping parameters, $t_z/t$, as a tuning parameter, which controls how corrugated the cylindrical-like Fermi surface is. We compute, within the random-phase-approximation (RPA), the real and imaginary parts of the susceptibility at the momentum $\mbf{Q}=(\pi,\pi,Q_z)$, and investigate the appearance of a spin-resonance mode for different $Q_z$ values. 

For $Q_z = 0$, we find two different scenarios. For sufficiently small $t_z/t$ (i.e. large anisotropy), a well-defined spin resonance mode emerges, separated from the particle-hole continuum by a sizeable gap in the magnetic spectrum. For these values of $t_z/t$, the ``hot lines'' formed by the Fermi surface points connected by $\mbf{Q}$ are open lines, like in the 2D case. When $t_z/t$ is increased, the effective excitation gap at the the open hot lines becomes non-uniform, being suppressed in certain regions. As a result, for weak interactions, the energy of the spin resonance mode lies above the particle-hole continuum, which makes it overdamped and incoherent. Conversely, for larger values of the interaction, the spin resonance mode remains sharp.
When $t_z/t$ is further increased above a threshold value, the hot lines close due to the corrugation of the Fermi surface, which remains open. Once the hot lines undergo this open-to-closed topological transition, they cross the superconducting nodal lines, causing the particle-hole continuum to extend all the way to zero-frequency. As a consequence, the resonance mode is overdamped and the associated maximum in the dynamical spin susceptibility becomes broad and incoherent, with no genuine resonance appearing at any frequency, regardless of the value of the interaction.Unlike the $Q_z = 0$ case, for $Q_z = \pi$ we find that the hot lines are always open and the spin resonance mode is preserved for the $t_z/t$ values studied. Therefore, in agreement with previous studies~\cite{chubukov2008spin}, our results show that the dispersion of the resonance peak is an important fingerprint of its microscopic origin.

This paper is organized as follows: To set the stage, we rederive the RPA dynamical spin susceptibility for a two-dimensional system with a nearly-nested Fermi surface in Sec.~\ref{sec:2d} and demonstrate how a sign-changing superconducting order parameter leads to a spin resonance mode. Sec.~\ref{sec:3d_cases} then presents our main results. In \ref{sec:susc} we introduce the model employed and study the corresponding static spin susceptibility in the superconducting phase. In \ref{sec:qz0} and \ref{sec:qzpi} we present numerical calculations of the dynamical spin susceptibility for magnetic scattering vectors $\mbf{Q}_M=(\pi,\pi,0)$ and $\mbf{Q}_A=(\pi,\pi,\pi)$. At $\mbf{Q}_M$ the appearance of a spin resonance mode is sensitive to the corrugation of the Fermi surface, while a robust spin resonance mode appears independent of the corrugation near $\mbf{Q}_A$. In Sec.~\ref{sec:0topi} we study the evolution of the spin resonance mode as a function of out-of-plane momentum between $\mbf{Q}_M$ and $\mbf{Q}_A$. Our conclusions are presented in Sec.~\ref{sec:Conclusions}.

\section{Spin resonance mode in two dimensions \label{sec:2d}}

In this section, as a warm-up, we re-derive the two-dimensional dynamical spin susceptibility in the case of a nearly nested Fermi surface inside a $d_{x^2-y^2}$-wave superconducting state. This section reproduces well-known results for the emergence of a spin-resonance mode in unconventional superconductors -- see for instance the reviews \onlinecite{Eschrig2006Effect,Abanov2003,scalapino2012}. Our starting point is the simple one-band Hubbard model with a superconducting pairing term,
\begin{equation}
\begin{aligned}H & =\sum_{\mathbf{k}\sigma}\xi_{\mathbf{k}}\hat{c}_{\mathbf{k\sigma}}^{\dagger}\hat{c}_{\mathbf{k\sigma}}+U\sum_{\text{\ensuremath{\mathbf{kk'q}}}}\hat{c}_{\mathbf{k+q\uparrow}}^{\dagger}\hat{c}_{\mathbf{k'-q\downarrow}}^{\dagger}\hat{c}_{\mathbf{k'\downarrow}}\hat{c}_{\mathbf{k\uparrow}}\\
 & \quad+\sum_{\mathbf{k}}\left(\Delta_{\mathbf{k}}\hat{c}_{\mathbf{k}\uparrow}^{\dagger}\hat{c}_{-\mathbf{k\downarrow}}^{\dagger}+\text{H.c.}\right)\,,
\end{aligned}
\label{eq:hamiltonian}
\end{equation}
where $\xi_{\mbf{k}}$ is the dispersion, $U>0$ is the Hubbard repulsion and $\Delta_{\mbf{k}}$ is the superconducting order parameter. The dynamical spin susceptibility is the retarded part of the spin-spin correlation function
\begin{equation}
    \chi_0^{ij}(\mathbf{q},\omega)=-i\int_0^{\infty}\mathrm{d}t e^{i(\omega + i 0^+)t}\left\langle S^{i}(\mathbf{q},t)S^{j}(-\mathbf{q},0)\right\rangle_0\,,\label{eq:bare_susc_GF}
\end{equation}
where $t > 0$ and the expectation value $\langle \cdot \rangle_0$ is taken with respect to the non-interacting part of the Hamiltonian which, in the superconducting phase, includes the second line of Eq.~\eqref{eq:hamiltonian}. The operator $S^{i}(\mbf{q},t)$ is given by
\begin{equation}
    S^{i}(\mathbf{q},t)=\sum_{\mathbf{k}\alpha\beta}c_{\mathbf{k+q}\alpha}^{\dagger}(t)\sigma_{\alpha\beta}^{i}c_{\mathbf{k}\beta}(t)\,.
\end{equation}
Since the spin-SU(2) symmetry remains unbroken in our model, we can focus on one component of the spin susceptibility and thus take $i=j=z$ without loss of generality. Hereafter, we will therefore suppress the spin-indices on the susceptibility. Using Wick's theorem, we obtain
\begin{align}
    \chi_0(\mbf{q},\omega) = \frac{i}{V}\sum_{\mbf{k}}\int \frac{\mathrm{d}\Omega}{2\pi} \big[ &G(\mbf{k}+\mbf{q},\Omega + \omega)G(\mbf{k},\Omega) \nonumber \\ + &F^{\dagger}(\mbf{k}+\mbf{q},\Omega + \omega)F(\mbf{k},\Omega) \big]\,,
    \label{eq:bare_spin_susc_freq_integral}
\end{align}
where $V$ is the $\mbf{k}$-space volume and
\begin{align}
    G(\mbf{k},\omega) &= \frac{\omega + i0^+ \text{sign}\omega + \xi_{\mbf{k}}}{(\omega + i 0^+ \text{sign}\omega)^2 -E_{\mbf{k}}^2}\,,\\
    F(\mbf{k},\omega) &= \frac{\Delta_{\mbf{k}}}{(\omega + i 0^+ \text{sign}\omega)^2 -E_{\mbf{k}}^2} 
\end{align}
are the normal and anomalous Green functions, respectively, with $E_{\mbf{k}}^2 = \xi_{\mbf{k}}^2 + \Delta_{\mbf{k}}^2$. As our interest is in the dynamical part of the susceptibility, we split up the bare susceptibility in a frequency-dependent, dynamic component and a frequency-independent, static component
\begin{equation}
    \chi_0 (\mbf{q},\omega) = \chi_0^{\rm stat}(\mbf{q}) + \chi_0^{\rm dyn}(\mbf{q},\omega)\,.
\end{equation}
In the remainder of this section we ignore the static contribution as it has no influence on the spin resonance mode.

\begin{figure}
\includegraphics[width=0.75\columnwidth]{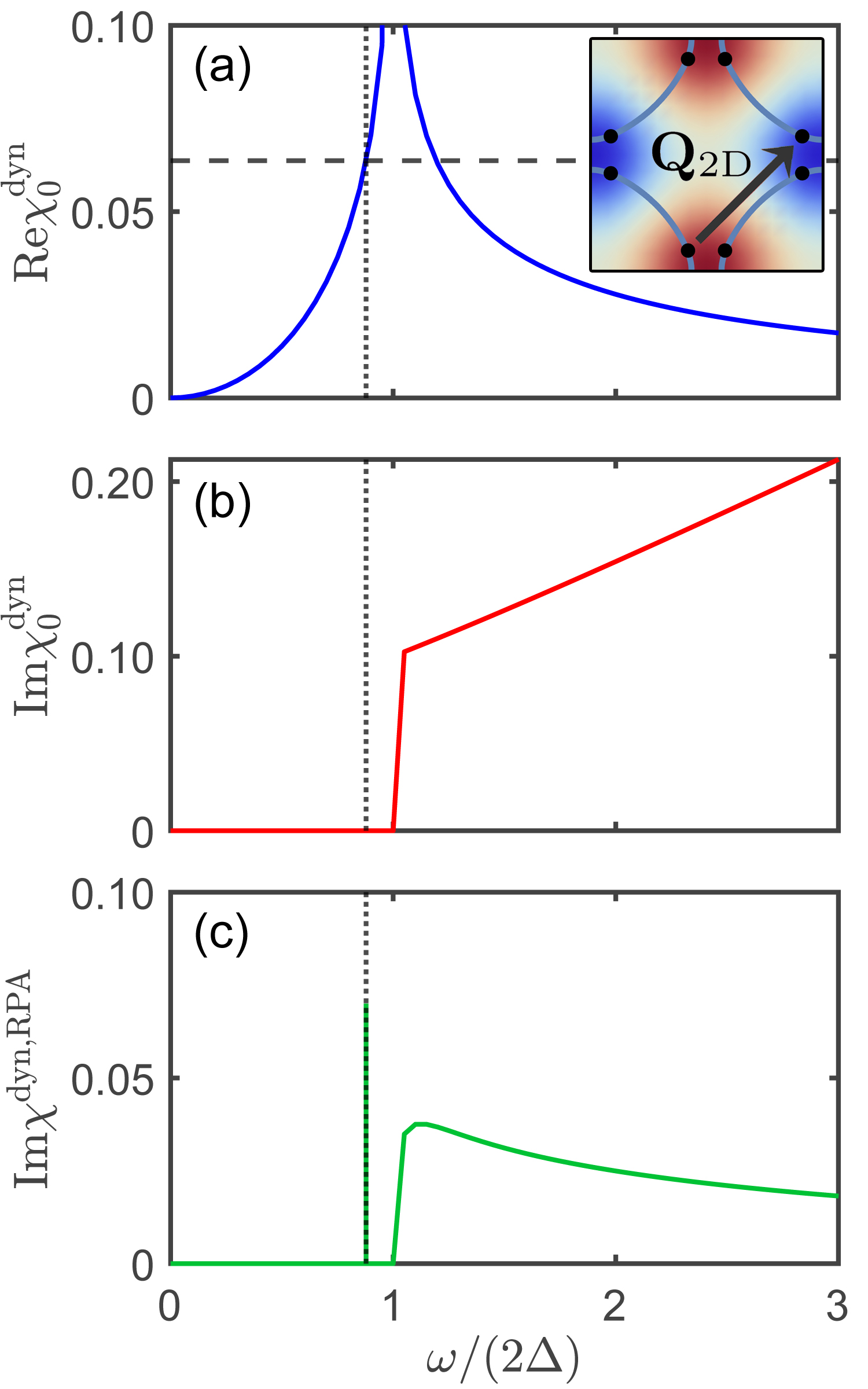}
\caption{(a) Real and (b) imaginary parts of the bare dynamical spin susceptibility, $\chi(\mbf{Q}_\text{2D},\omega)$, in arbitrary units, at the antiferromagnetic wave-vector $\mbf{Q_{\rm 2D}}=(\pi,\pi)$ and for a sign-changing superconducting order parameter. The real part exhibits a logarithmic divergence at $\omega=2\Delta$ while the imaginary part vanishes for $\omega < 2\Delta$. (c) Imaginary part of the RPA spin susceptibility $\text{Im} \chi^{\rm RPA}(\mbf{Q}_\text{2D},\omega)$, for the value of $U^{-1}$ denoted by the black dashed line in (a). There is a clear resonance peak in $\text{Im} \chi^{\rm RPA}(\mbf{Q}_\text{2D},\omega)$ at the frequency where $\text{Re}\chi_0(\mbf{Q}_\text{2D},\omega_c)=1/U$, $\text{Im} \chi_0(\mbf{Q}_\text{2D},\omega_c)= 0$, which is below $2\Delta$. The inset shows the Fermi surface with the antiferromagnetic wave-vector (black arrow) and hot spots (black dots). The underlying color scale denotes the sign and magnitude of the superconducting order parameter.}
\label{fig:2danalytic}
\end{figure}

The main contribution to the spin resonance mode in the current case arises from hot spots of the Fermi surface, i.e. the points connected by the antiferromagnetic wave-vector $\mbf{Q}_\text{2D}=(\pi,\pi)$ , shown in the inset of Fig.~\ref{fig:2danalytic}(a). Focusing near this wave-vector, we set $\mbf{q}=\mbf{Q}_\text{2D}$ in Eq.~\eqref{eq:bare_spin_susc_freq_integral} and linearize the dispersion near the Fermi level, such that $\xi_{\mbf{k}} \approx \mbf{v}\cdot (\mbf{k}-k_F)$. Because the superconducting order parameter has $d_{x^2-y^2}$ symmetry, it changes sign between the hot spots connected by $\mbf{Q}_\text{2D}$: $\Delta_{\mbf{k}}=-\Delta_{\mbf{k}+\mbf{Q}_\text{2D}}=\Delta$. Using these simplifications, we can integrate over momentum in Eq.~\eqref{eq:bare_spin_susc_freq_integral} and obtain~\cite{chubukov2008spin}
\begin{equation}
    \chi_{0}^{\rm dyn}(\mbf{Q}_\text{2D},\omega) \propto i\int\frac{d\Omega}{2\pi}\left(1-\frac{\Omega_{+}\Omega_{-}-\Delta^{2}}{\sqrt{\Omega_{+}^{2}-\Delta^{2}}\sqrt{\Omega_{-}^{2}-\Delta^{2}}}\right)\,, \label{eq:2dexp}
\end{equation}
where $\Omega_{\pm}=\Omega\pm\frac{\omega}{2}+i0^+\text{sign}(\Omega\pm\frac{\omega}{2})$ and the proportionality includes a factor dependent on the relative Fermi velocities at the hot spots. As widely discussed in the literature  \cite{Eschrig2006Effect,Abanov2003,scalapino2012}, the minus sign in the numerator arises due to the sign-change in the gap function and is crucial for the emergence of the spin resonance mode.

The real and imaginary parts of $\chi_0^{\rm dyn}(\mbf{Q}_\text{2D},\omega)$ are shown in Figs.~\ref{fig:2danalytic}(a) and \ref{fig:2danalytic}(b).
The real part exhibits a logarithmic divergence at $\omega= 2\Delta$, which, \emph{via} the Kramers-Kronig relations, implies a jump in the imaginary part, which vanishes for $\omega < 2\Delta$ and increases linearly after the jump \cite{Abanov2003}. The dynamic susceptibility only exhibits a pole once interactions are included. This can be achieved within RPA, which gives \cite{Eschrig2006Effect}
\begin{equation}
    \chi^{\rm RPA}(\mbf{q},\omega) = \frac{\chi_0(\mbf{q},\omega)}{1-U\chi_0 (\mbf{q},\omega)}\,. \label{eq:RPA}
\end{equation}
From this expression it is evident that a pole appears when $1- U \chi_0 (\mbf{q},\omega)=0$. As $\chi_0(\mbf{q},\omega)$ is a complex function [see Figs.~\ref{fig:2danalytic}(a) and (b)] the condition for the existence of a pole becomes $1-U \text{Re}\chi_0(\mbf{q},\omega)=0$ and $\text{Im}\chi_0(\mbf{q},\omega)=0$. As shown in Fig.~\ref{fig:2danalytic}(a) the real part of the susceptibility at the wavevector $\mbf{q}=\mbf{Q}_\text{2D}$ exhibits a divergence at $\omega = 2\Delta$, while the imaginary part vanishes for $\omega < 2\Delta$. Hence, an infinitesimal $U$ is sufficient to lead to the appearance of a pole in the RPA susceptibility whose frequency decreases with increasing $U$ \cite{Norman2000,Abanov2003,Eremin2008Feedback,Eschrig2006Effect,scalapino2012}. In Fig.~\ref{fig:2danalytic}(c) we show the RPA susceptibility obtained for a given $U$ [black dashed line in Fig.~\ref{fig:2danalytic}(a)]; a resonance mode is clearly observed at $U=1/\text{Re}\chi_0(\mbf{Q}_\text{2D},\omega_c)$. Ultimately, if $U$ reaches the critical value, $U_c = 1/N(\epsilon_F)$, where $N(\epsilon_F)$ is the density of states at the Fermi level, the mode condenses, and a magnetically ordered state emerges \cite{Ismer2007,Rosa2017}. Note that, since Fig.~\ref{fig:2danalytic}(a) only shows the dynamical part of the susceptibility, this fact is not apparent from the figure. If the imaginary part of the susceptibility attains a finite value, e.g., as a consequence of temperature, the pole in the RPA susceptibility is replaced by a broadened peak and the associated mode acquires a finite lifetime.

The above discussion focuses on the impact of a sign-changing superconducting order parameter in two dimensions. The gap in the imaginary part of the bare susceptibility alongside the divergent nature of the real part ensures the presence of a well-defined spin resonance mode. This is in stark contrast to the case of $s$-wave superconductivity, which features no spin resonance mode~\cite{scalapino2012}.

\section{Spin resonance mode in three-dimensional systems}\label{sec:3d_cases}

The dynamical spin susceptibility of an isotropic 3D unconventional superconductor with a spherical Fermi surface was analyzed in Ref.~\onlinecite{chubukov2008spin}. The main result was that, in contrast to the well-known 2D case, the spin-resonance mode ceased to exist. Our goal is to elucidate the mechanisms responsible for the suppression of the spin resonance as the system smoothly evolves from quasi-2D to more 3D-like. To do that, we consider a generic anisotropic three-dimensional system with a corrugated, nearly cylindrical Fermi surface. In our approach, the hopping parameter along the $\hat{z}$-direction remains a fraction of the in-plane hopping parameter, such that the Fermi surface itself remains open and nearly cylindrical. This choice is motivated by the fact that many unconventional superconductors are layered materials, exhibiting significant anisotropy between in- and out-of-plane directions. We restrict attention to the usual $d_{x^2-y^2}$-wave form factor and focus on how the corrugation of the Fermi surface impacts the appearance of a resonance peak.

\subsection{Microscopic model and spin susceptibility}\label{sec:susc}

We depart from the one-band Hubbard Hamiltonian introduced in Eq.~\eqref{eq:hamiltonian}, albeit with the momentum sums being over the 3D Brillouin zone and the dispersion taking the form:
\begin{align}
    \xi_{\mbf{k}} = & 2t \left( \cos k_x a + \cos k_y a \right) -4 t' \cos k_x a \cos k_y a \nonumber \\ + & 2 t_z \cos k_z a - \mu\,.\label{eq:3d_dispersion}
\end{align}
\begin{figure}
\includegraphics[width=0.8\columnwidth]{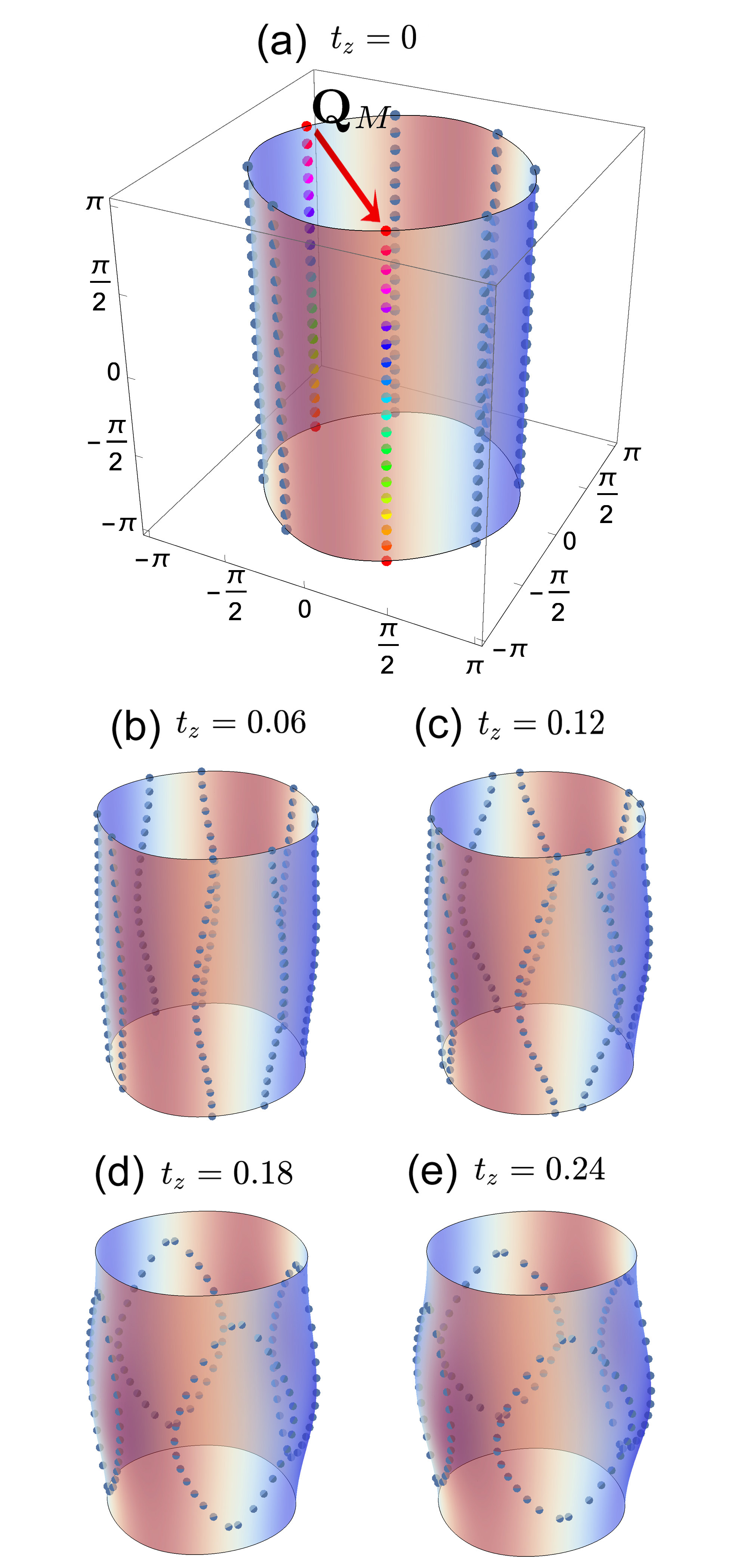}
\caption{Fermi surfaces for the dispersion in Eq.~\eqref{eq:3d_dispersion} with $t=1$,  $t'=-0.25$, $n=1.10$, and increasing values of $t_z$ as denoted in each panel. The color scale denotes the sign and magnitude of the superconducting gap. The dots along the Fermi surfaces denote the hot lines connected by the in-plane antiferromagnetic wave-vector $\mbf{Q}_M=(\pi,\pi,0)$. In (a), the pairs of spots along the hot lines connected by $\mbf{Q}_M$ are highlighted with the same color. As $t_z$ is increased from (a) $t_z=0$ to (e) $t_z=0.24$, the corrugation of the Fermi surface changes, which causes a topological transition in the shape of the hot lines. For $t_z=0$ [panel (a)], $t_z = 0.06$ [panel (b)] and $t_z=0.12$ [panel (c)] the hot lines are open, while for $t_z=0.18$ [panel (d)] and $t_z=0.24$ [panel (e)] the hot lines form closed loops that intersect the superconducting nodal lines.}
\label{fig:fs}
\end{figure}
The superconducting order parameter is assumed to be given by the $d_{x^2-y^2}$-wave gap 
\begin{equation}
    \Delta_{\mbf{k}} = \Delta_0 \left( \cos k_x a - \cos k_y a \right)\,,\label{eq:gap_form}
\end{equation}
which is consistent with the results of a weak-coupling study of the superconducting instabilities of the 3D anisotropic Hubbard model~\cite{Roising2018weak}.
The gap thus remains independent of $k_z$ and displays nodal lines described by $k_x = \pm k_y$. We will measure length in units of the lattice spacing $a=1$ and energy in units of the nearest-neighbor hopping $t=1$. The next-nearest neighbor in-plane hopping parameter is set to $t'=-0.25$, and the gap, to $\Delta_0=0.1$. The out-of-plane hopping parameter, $t_z$, is varied between $t_z=0$ and $t_z=0.24$, resulting in the increasingly corrugated cylindrical-like Fermi surfaces shown in Fig.~\ref{fig:fs}, which remains open. The chemical potential, $\mu$, is chosen to keep the electron density in the normal state, $n=1.10$, fixed between the various cases. As we argue below, the specific choice of parameters does not alter our main conclusions.

\begin{figure}
\includegraphics[width=1\columnwidth]{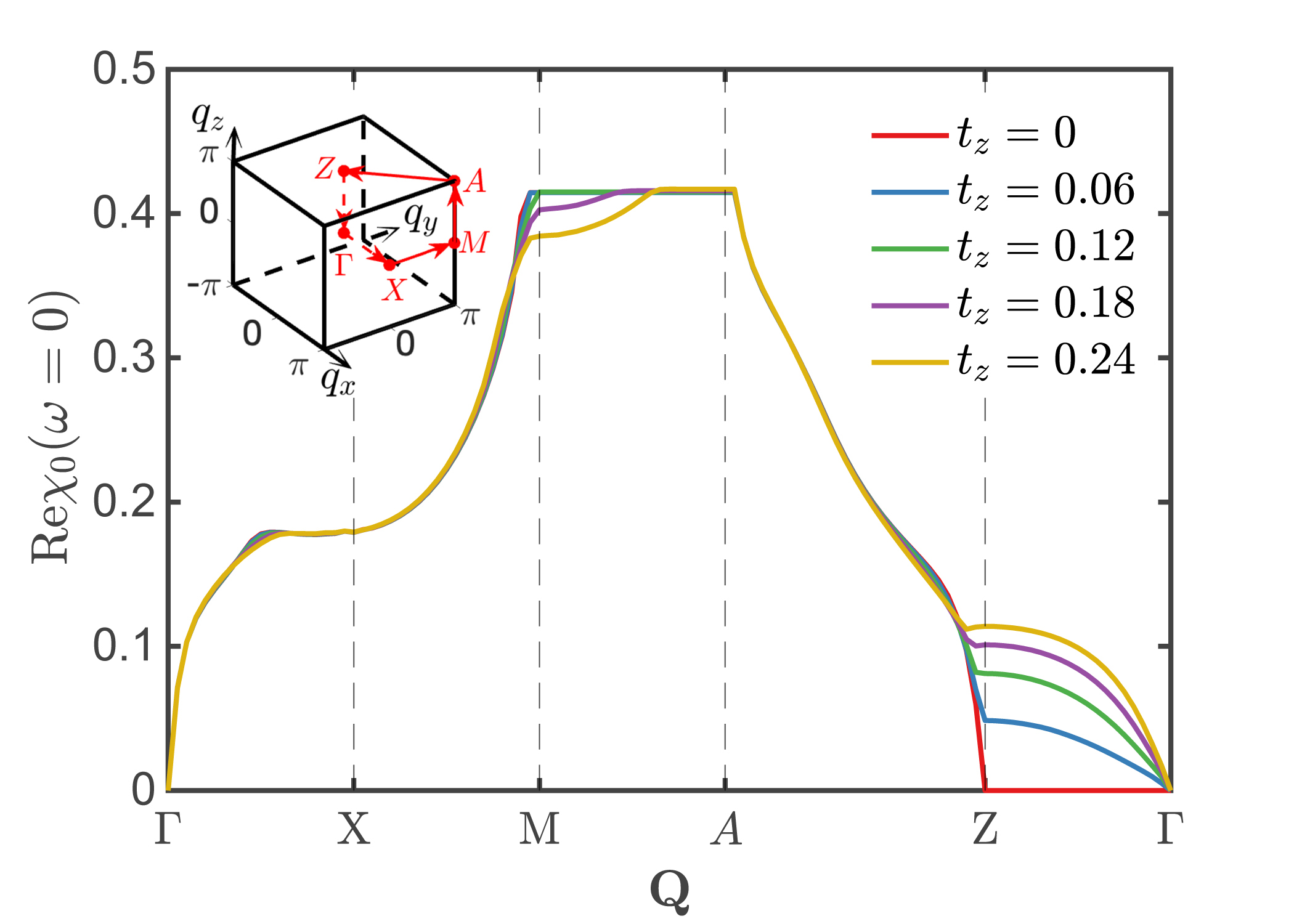}
\caption{Real part of the bare static spin susceptibility (in arbitrary units) in the superconducting phase, $\text{Re}\chi_{0}(\mbf{q},\omega=0)$, along a high-symmetry path in the three-dimensional Brillouin zone (shown  in the inset) for different values of $t_z$. Regardless of the value of $t_z$, the spin susceptibilities are clearly peaked along the line $M=(\pi,\pi,0)$ to $A=(\pi,\pi,\pi)$. While $t_z$ weakly suppresses the peak at $M$, its main effect is along the path from $\Gamma=(0,0,0)$ to $Z=(0,0,\pi)$.}
\label{fig:chi0w0}
\end{figure}

The general expression for the non-interacting spin susceptibility in the superconducting phase is given in Eq.~\eqref{eq:bare_spin_susc_freq_integral}. The discussion in Sec.~\ref{sec:2d} centered on a specific value for the momentum transfer $\mbf{q}$, which allowed us to carry out the momentum integration analytically. For a generic $\mbf{q}$, we can instead carry out the frequency integral in Eq.~\eqref{eq:bare_spin_susc_freq_integral}, obtaining~\cite{bulut1993}
\begin{align}
    & \chi_{0}(\mathbf{q},\omega)=-\frac{1}{V}\sum_{\mathbf{k}}\nonumber\\
 & \left\{ \frac{1}{2}\left[1+\frac{\xi_{\mathbf{k}}\xi_{\mathbf{k+q}}+\Delta_{\mathbf{k+q}}\Delta_{\mathbf{k}}}{E_{\mathbf{k}}E_{\mathbf{k+q}}}\right]\frac{n_{F}(E_{\mathbf{k}})-n_{F}(E_{\mathbf{k+q}})}{\omega+E_{\mathbf{k}}-E_{\mathbf{k+q}}+i\Gamma}\right. \nonumber \\
 & +\frac{1}{4}\left[1-\frac{\xi_{\mathbf{k}}\xi_{\mathbf{k+q}}+\Delta_{\mathbf{k+q}}\Delta_{\mathbf{k}}}{E_{\mathbf{k}}E_{\mathbf{k+q}}}\right]\frac{n_{F}(E_{\mathbf{k}})+n_{F}(E_{\mathbf{k+q}})-1}{\omega+E_{\mathbf{k}}+E_{\mathbf{k+q}}+i\Gamma} \nonumber \\
 & \left.+\frac{1}{4}\left[1-\frac{\xi_{\mathbf{k}}\xi_{\mathbf{k+q}}+\Delta_{\mathbf{k+q}}\Delta_{\mathbf{k}}}{E_{\mathbf{k}}E_{\mathbf{k+q}}}\right]\frac{1-n_{F}(E_{\mathbf{k}})-n_{F}(E_{\mathbf{k+q}})}{\omega-E_{\mathbf{k}}-E_{\mathbf{k+q}}+i\Gamma}\right\} \label{eq:3dchi0}
\end{align}
where $n_F(\cdot)$ denotes the Fermi-Dirac distribution function and $\Gamma$ is a lifetime parameter introduced to mimic thermal broadening and ensure convergence of the numerical momentum summations. Hereafter, we set $k_{\rm B}T=0.01$ and $\Gamma=0.01$. We refer to $\chi_{0}$ as the \textit{bare susceptibility} inside the superconducting state; the RPA susceptibility is calculated according to Eq. (\ref{eq:RPA}). We evaluate the momentum summation in Eq.~\eqref{eq:3dchi0} numerically using a uniform momentum-grid with $2000^3$ points. This is done to ensure accurate results on the frequency-axis, which is more sensitive to the number of points used in momentum space. The static spin susceptibility, $\omega=0$, is shown in Fig.~\ref{fig:chi0w0} for $\mbf{q}$ along a high-symmetry path of the Brillouin zone (shown in the inset) and for different values of $t_z$, for which the Fermi surfaces are shown in Fig.~\ref{fig:fs}.

The main impact of $t_z$ is seen along the direction from $\Gamma=(0,0,0)$ to $Z=(0,0,\pi)$. For $t_z=0$, the system is effectively two-dimensional and no variation is found along this direction. Increasing $t_z$ adds a dispersion along the $k_z$-direction and alters the susceptibility. The prominent peaks at $M=(\pi,\pi,0)$ and $A=(\pi,\pi,\pi)$ are a direct consequence of the nested character of the Fermi surfaces of Fig.~\ref{fig:fs}. The corrugation of the Fermi surface at finite values of $t_z$ acts to suppress nesting by $(\pi,\pi,0)$ slightly, and thus diminishes the associated peak. Motivated by the shape of the static susceptibility in the superconducting state, in what follows we will focus on two distinct antiferromagnetic wave-vectors, $\mbf{Q}_M=(\pi,\pi,0)$ and $\mbf{Q}_A=(\pi,\pi,\pi)$, and contrast the behavior of the dynamical spin susceptibility near these two momenta.

\subsection{Dynamic susceptibility at $\mbf{Q}_M=(\pi,\pi,0)$}\label{sec:qz0}

The wave-vector $\mbf{Q}_M=(\pi,\pi,0)$ is illustrated as the vector connecting identically colored spots in Fig.~\ref{fig:fs} for $t_z=0$. This results in so-called hot lines, which generalize the hot spots appearing in two dimensions. As $t_z$ is increased, the hot lines undergo a topological transition from open to closed. The closed hot-line loops cross the superconducting line nodes, as seen in Figs.~\ref{fig:fs}(d) and (e). Note that, as in the 2D case, $\mbf{Q}_M$ connects regions of the Fermi surface for which the gap has opposite signs. 

\begin{figure}
\includegraphics[width=1\columnwidth]{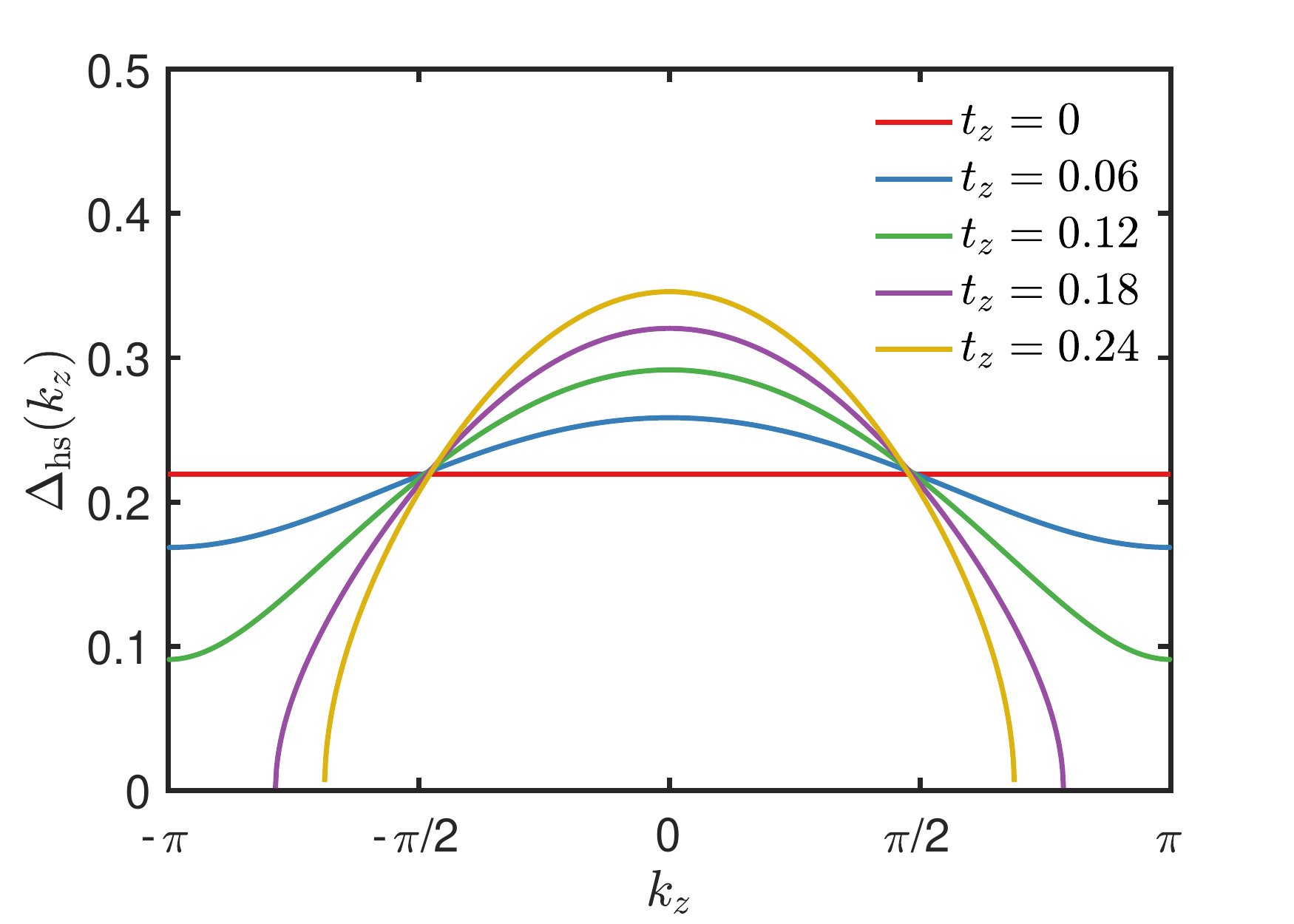}
\caption{Effective excitation gap on the hot spots $\Delta_{\text{hs}}(k_z) = \abs{\Delta_\mbf{k}}+ \abs{\Delta_\mbf{k+Q}}$ measured along the hot lines in Fig.~\ref{fig:fs} for different values of $t_z$. For $t_z=0.18$ and $t_z=0.24$, where the hot lines form closed loops, $\Delta_{\text{hs}}(k_z)$ goes to zero at the value of $k_z$ for which the loop crosses the superconducting nodal line. Beyond this value of $k_z$, $\Delta_{\text{hs}}(k_z)$ is not defined.}
\label{fig:hl1}
\end{figure}

In Fig.~\ref{fig:hl1} we show the behavior of the effective excitation gap at the hot spots, $\Delta_{\text{hs}}(k_z) = \abs{\Delta_\mbf{k}}+ \abs{\Delta_\mbf{k+Q}}$, for different values of $t_z$. This quantity plays a similar role as $2\Delta$ in the 2D case, with the main difference being its $k_z$ momentum dependence. This defines two different characteristic excitation gaps. $\Delta_{\rm exc}=\min \left[ \Delta_{\text{hs}}(k_z) \right]$ denotes the minimum energy cost for excitations with momentum $\mbf{Q}_M$. Conversely, $\Delta_{\rm max} \equiv \max \left[ \Delta_{\text{hs}}(k_z) \right]$ denotes the maximum excitation gap along the hot lines. For $t_z=0.18$ and $t_z=0.24$, beyond the threshold $t_z$ value for which the hot lines undergo a topological transition, $\Delta_{\text{hs}}(k_z)$ is only defined on a limited range of $k_z$ values, since the hot lines form closed loops on the Fermi surface. Importantly, $\Delta_{\rm exc}=0$ for the closed loops, since they cross the superconducting nodal lines defined by $k_x = \pm k_y$.
\begin{figure}
	\includegraphics[width=1\columnwidth]{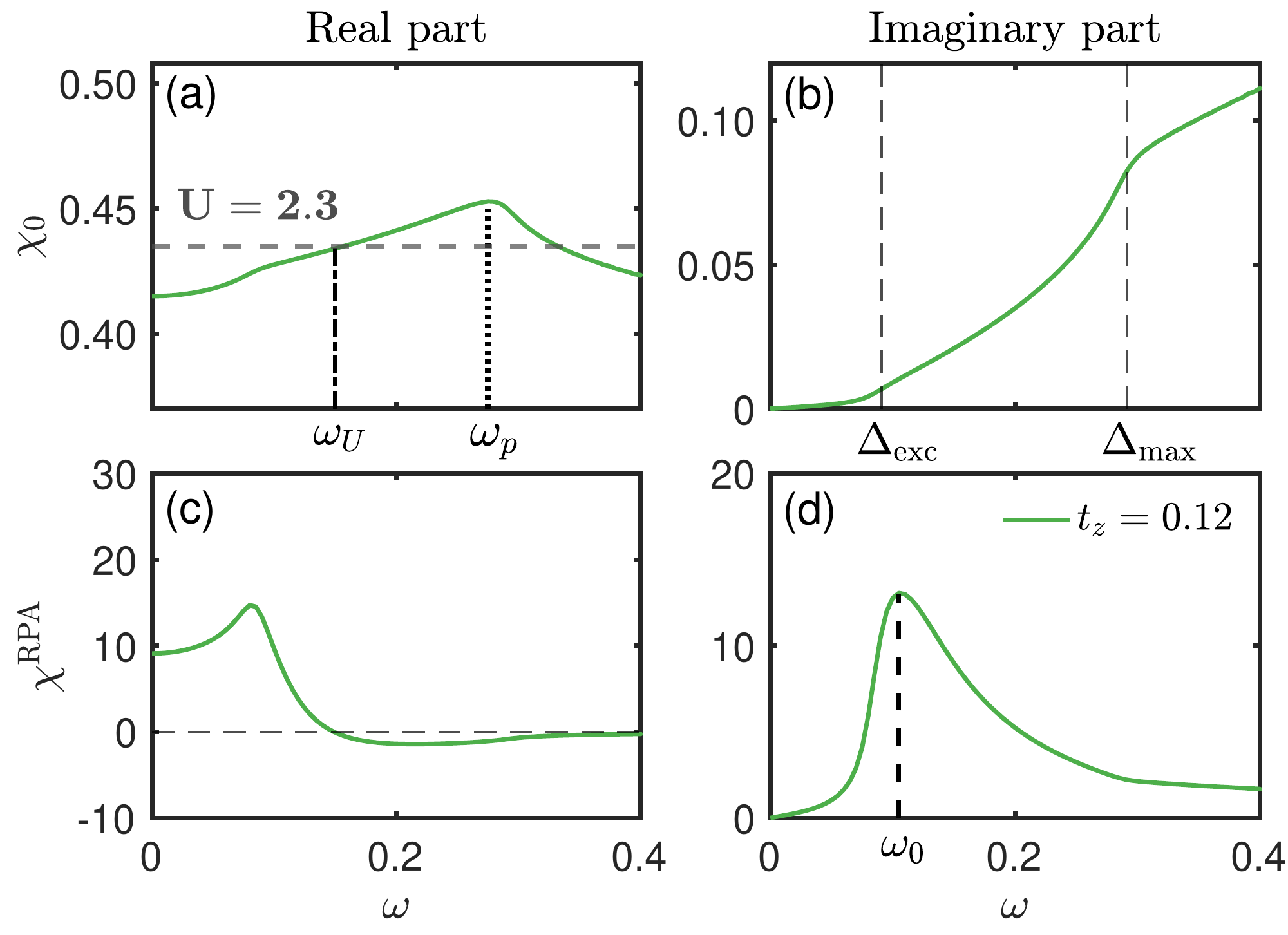}
	\caption{Real (left column) and imaginary (right column) parts of the bare (upper row) and RPA (lower row) spin susceptibilities at $\mbf{Q}_M$ for $t_z =0.12$. In (a), $\omega_p$ (dashed) is defined as the frequency where $\text{Re}\chi_0(\mbf{Q}_M,\omega)$ reaches a maximum, whereas $\omega_U$ (dashed-dotted) is defined as the (lowest) frequency where $\text{Re}\chi_0(\mbf{Q},\omega_U)=1/U$. In (b), $\Delta_{\rm exc}$ and $\Delta_{\rm max}$ for this particular value of $t_z$ are denoted. We note that $\omega_p \approx \Delta_{\rm max}$, which holds regardless of the value of $t_z$. In (d), $\omega_0$ is defined as the frequency at which the imaginary part of the RPA susceptibility is peaked. Note that both $\omega_U$ and $\omega_0$ depend on $U$.}
\label{fig:wp}
\end{figure}

\begin{figure}
\includegraphics[width=1\columnwidth]{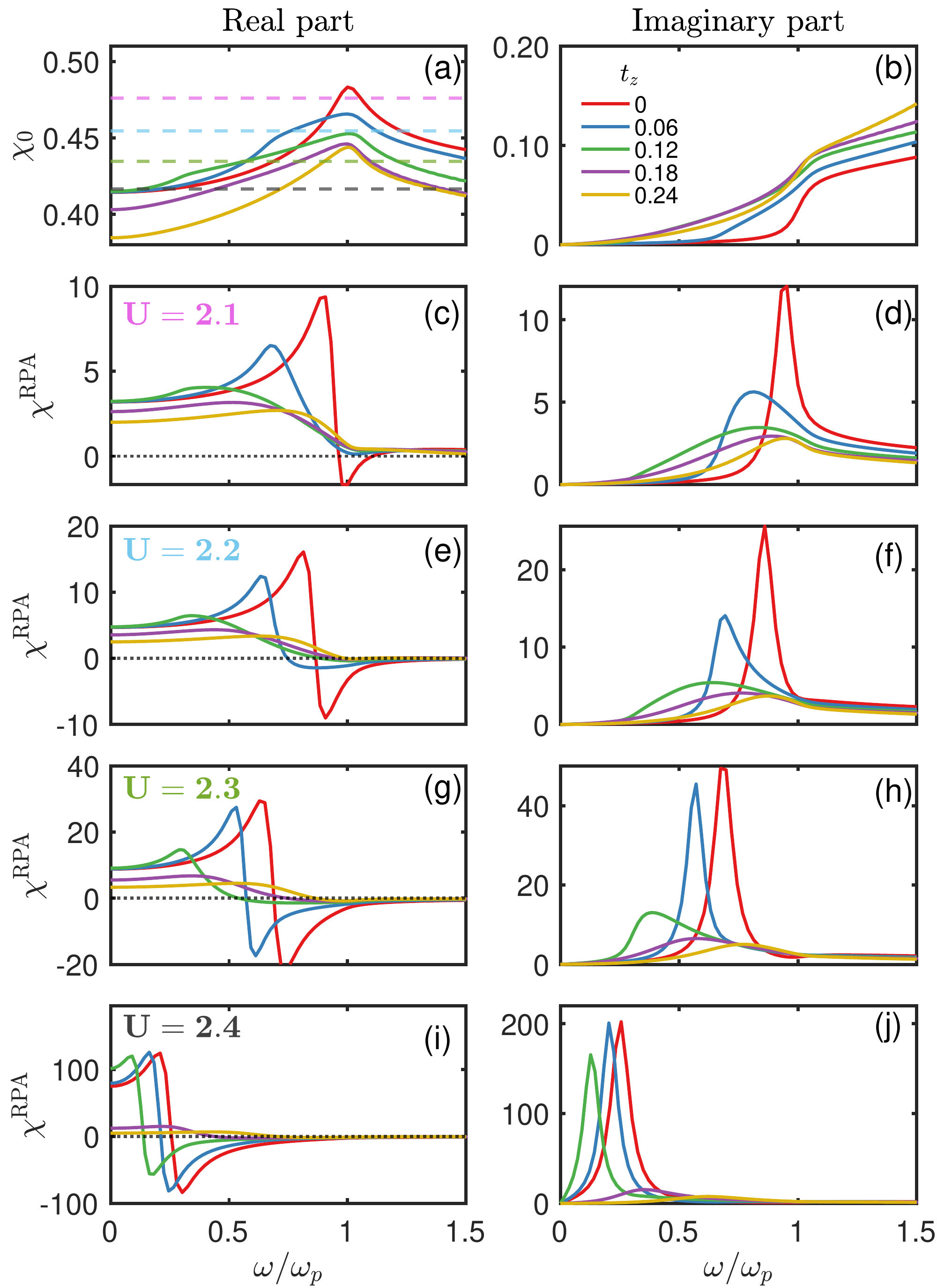}
\caption{Real (left column) and imaginary (right column) parts of the spin susceptibility at $\mbf{Q}_M=(\pi,\pi,0)$ for different values of $U$. The different colors correspond to different values of $t_z$, as indicated in the legend. Here, the $x$-axis has been rescaled by $\omega_p$, defined as the frequency at which $\text{Re}\chi_0(\mbf{Q}_M,\omega)$ has a maximum (see Fig.~\ref{fig:wp}). In (a), we show the real part of the bare susceptibility ($U=0$), $\text{Re}\chi_0(\mbf{Q}_M,\omega)$, while the dashed horizontal lines denote $1/U$ for progressively larger values of $U$, for which the RPA susceptibility is plotted in (c)--(j). The imaginary part of the bare susceptibility, $\text{Im}\chi_0(\mbf{Q}_M,\omega)$, is shown in (b). Evidently, as $t_z$ is increased, the particle-hole continuum extends to lower and lower frequencies.}
\label{fig:chi0rpa}
\end{figure}

\begin{figure}
\includegraphics[width=1\columnwidth]{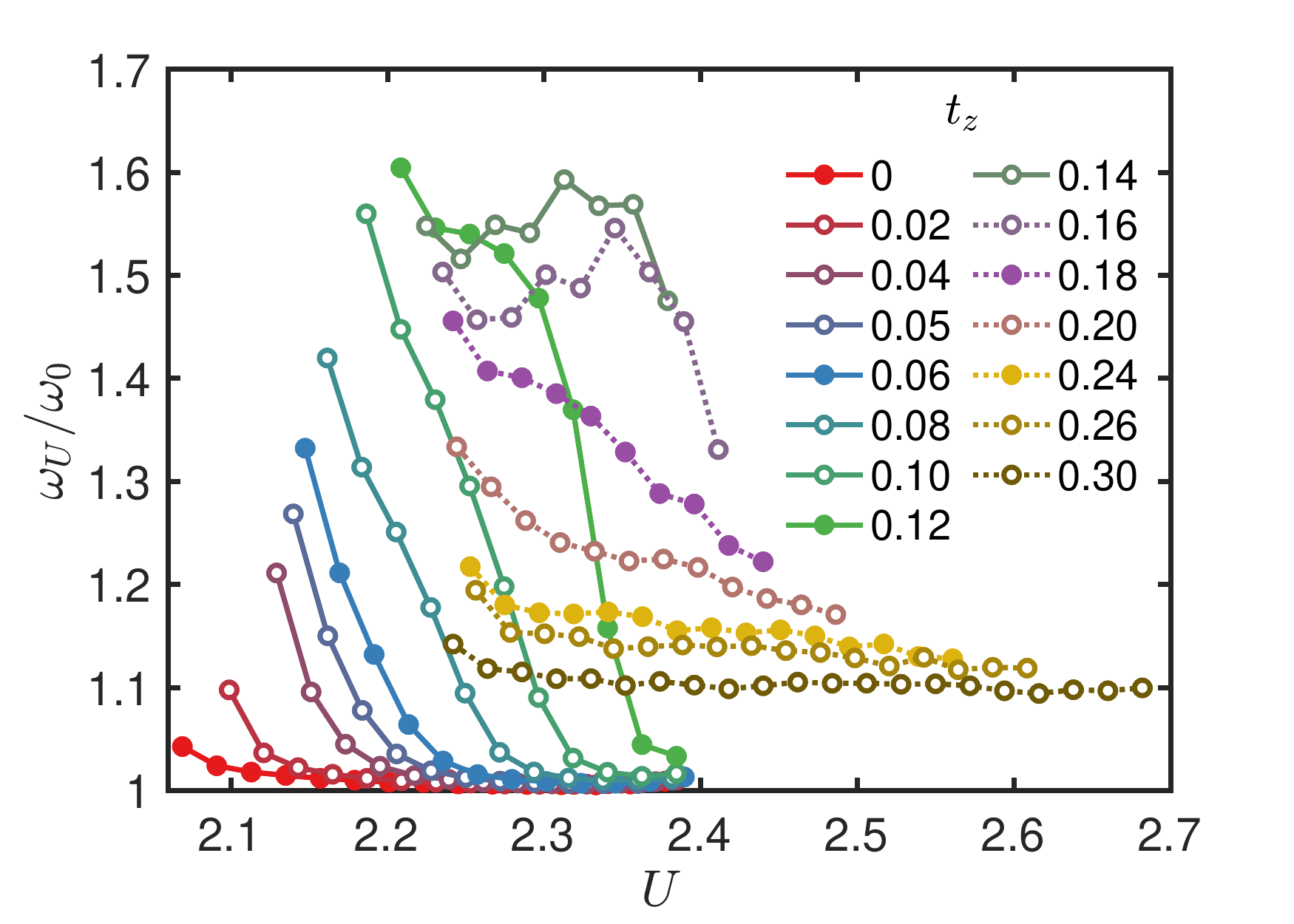}
\caption{Ratio of the characteristic frequencies $\omega_U$ and $\omega_0$ as a function of $U$ for different values of $t_z$. Full (dotted) lines correspond to $t_z$ values with open (closed) hot lines. The lines with full symbols are those studied in additional detail in Figs.~\ref{fig:fs}--\ref{fig:chi0rpa}. In the cases exhibiting a spin resonance, the ratio tends to one, whereas in the cases where no spin resonance appears, the ratio remains larger than one.}
\label{fig:freq_ratio}
\end{figure}

The hot-line features described above are manifested in the real and imaginary parts of the bare dynamic susceptibility at the wavevector $\mbf{Q}_M$ shown in Figs.~\ref{fig:wp}(a) and (b), as well as in the RPA susceptibility shown in Figs.~\ref{fig:wp}(c) and (d) for $t_z=0.12$. The peak in the real part of the bare susceptibility occurs at a frequency set by the maximum excitation gap along the hot lines, i.e. $\omega_p \approx \Delta_{\rm max}$ [Fig.~\ref{fig:wp}(a)]. Concomitantly, the imaginary part of the bare susceptibility experiences a suppression of spectral weight [Fig.~\ref{fig:wp}(b)], but not a complete depletion, which only happens below $\Delta_{\rm exc}$. Note that the exponentially small spectral weight below $\Delta_{\rm exc}$ in Fig.~\ref{fig:wp}(b) is a consequence of the finite temperature and broadening adopted in the calculations. At zero temperature, the spectral weight would be completely depleted as in the 2D case shown in Fig.~\ref{fig:2danalytic}(b). We note that the relation $\omega_p \approx \Delta_{\rm max}$ holds independently of $t_z$, and $\omega_p$ therefore increases with increasing $t_z$.

For a given value of $U$, it is helpful to define another frequency scale $\omega_U$ as the (lowest) frequency for which $\text{Re}\chi_0(\mbf{Q},\omega_U)=1/U$, as illustrated in Fig.~\ref{fig:wp}(a). This is precisely one of the conditions for the emergence of a resonance mode. The second condition (i.e. vanishing of the imaginary part of the bare susceptibility) is only satisfied if $\omega_U < \Delta_{\rm exc}$. In this case, a proper spin resonance mode emerges. Otherwise, if $\omega_U > \Delta_{\rm exc}$, the mode decays into the particle-hole continuum, and a proper resonance mode is not present. This is precisely the case shown in Fig.~\ref{fig:wp}(d) for $t_z=0.12$ and $U=2.3$; instead of a sharp resonance, only a broad peak at a frequency $\omega_0$ appears in the imaginary part of the RPA susceptibility. As $U$ is increased, $\omega_U$ decreases and eventually becomes smaller than $\Delta_{\rm exc}$. In this case, there is a gap preventing the decay of the mode into the particle-hole continuum. As a result, a resonance mode can appear in the imaginary part of RPA spin susceptibility.

In Fig.~\ref{fig:chi0rpa} we show the real and imaginary parts of the susceptibility for increasing values of $U$ and $t_z$. In this figure, we rescaled the frequency-axis by $\omega_p$. Except for panels (a) and (b), the RPA susceptibility is shown. The logarithmic divergence of $\text{Re}\chi_0(\mbf{Q},\omega)$ seen in the 2D case in Fig.~\ref{fig:2danalytic}(a) is cut-off in this case due to the finite temperature and lifetime, which implies the absence of a discontinuity in $\text{Im}\chi_0(\mbf{Q}_M,\omega)$ in Fig.~\ref{fig:2danalytic}(b). Figs.~\ref{fig:chi0rpa}(c)--(j) depict the real and imaginary parts of $\chi^{\rm RPA}(\mbf{Q}_M,\omega)$ for progressively larger values of $U$. Based on the intuition developed above from the analysis of Fig.~\ref{fig:wp}, we can understand the different regimes that emerge. Recall that, in the two-dimensional case, a resonance mode always appears for values of $U$ for which $\text{Re}\chi_0(\mbf{Q}_M,\omega)=1/U$. In Fig.~\ref{fig:chi0rpa}(a) the dashed horizontal lines denote $1/U$. We choose $U$ values below the critical $U_c$ that would lead to a magnetic instability ($U_c \approx 2.41$ for $t_z=0$). For $t_z=0$, $\Delta_{\rm exc} = \Delta_{\rm max}$ (see Fig.~\ref{fig:hl1}), and a spin resonance mode can appear for any $U$ for which the equation $\text{Re}\chi_0(\mbf{Q},\omega)=1/U$ has a solution, as evidenced, e.g., in Fig.~\ref{fig:chi0rpa}(d). However, for finite values of $t_z$, the excitation gap varies along the hot lines and $\Delta_{\rm exc} < \Delta_{\rm max}$. Hence, as explained above, it is no longer sufficient that $\text{Re}\chi_0(\mbf{Q},\omega)=1/U$ has a solution for a spin resonance mode to appear. In this case, we must additionally require that the solution is present for frequencies $\omega < \Delta_{\rm exc}$, i.e. $\text{Re}\chi_0\left(\mbf{Q},\omega < \Delta_{\rm exc}\right)=1/U$ must have a solution. This is the scenario for $t_z=0.06$ and $t_z=0.12$: while a sharp resonance mode is absent for smaller values of $U$ [Figs.~\ref{fig:chi0rpa}(d) and (f)], it is present for larger $U$ values [Fig.~\ref{fig:chi0rpa}(h) for $t_z = 0.06$ and Fig.~\ref{fig:chi0rpa}(j) for $t_z = 0.12$]. Therefore, unlike the 2D case, even in a 3D system with a relatively large anisotropy, a resonance mode is no longer guaranteed to appear at weak coupling. Evidently, if the minimum excitation gap goes to zero along the hot lines, $\Delta_{\rm exc} = 0$, the requirement $\text{Re}\chi_0\left(\mbf{Q},\omega < \Delta_{\rm exc}\right)=1/U$ cannot be satisfied and no spin resonance mode will appear. This is the scenario for the $t_z=0.18$ and $t_z=0.24$ cases, in which no peak appears in the RPA susceptibility even at high values of $U$, as shown in Fig.~\ref{fig:chi0rpa}(j).

Based on these considerations, there are two distinct situations, based on the presence or absence of a spin resonance mode. To qualitatively distinguish these cases, we consider the ratio between two $U$-dependent frequencies $\omega_U$ and $\omega_0$ introduced in Fig.~\ref{fig:wp}. From the discussion in the preceding paragraph, we conclude that a spin resonance mode appears in the cases where the ratio between these frequencies tends to one, $\frac{\omega_U}{\omega_0}\approx 1$. This ratio is shown in Fig.~\ref{fig:freq_ratio} as a function of $U$ for different values of $t_z$ between $t_z=0$ and $t_z=0.30$. There is a clear qualitative distinction between the cases exhibiting spin resonance modes (plotted in solid lines) and the cases that do not (shown with dotted lines).

\subsection{Dynamic susceptibility at $\mbf{Q}_A=(\pi,\pi,\pi)$ \label{sec:qzpi}}

\begin{figure*}
\includegraphics[width=1\textwidth]{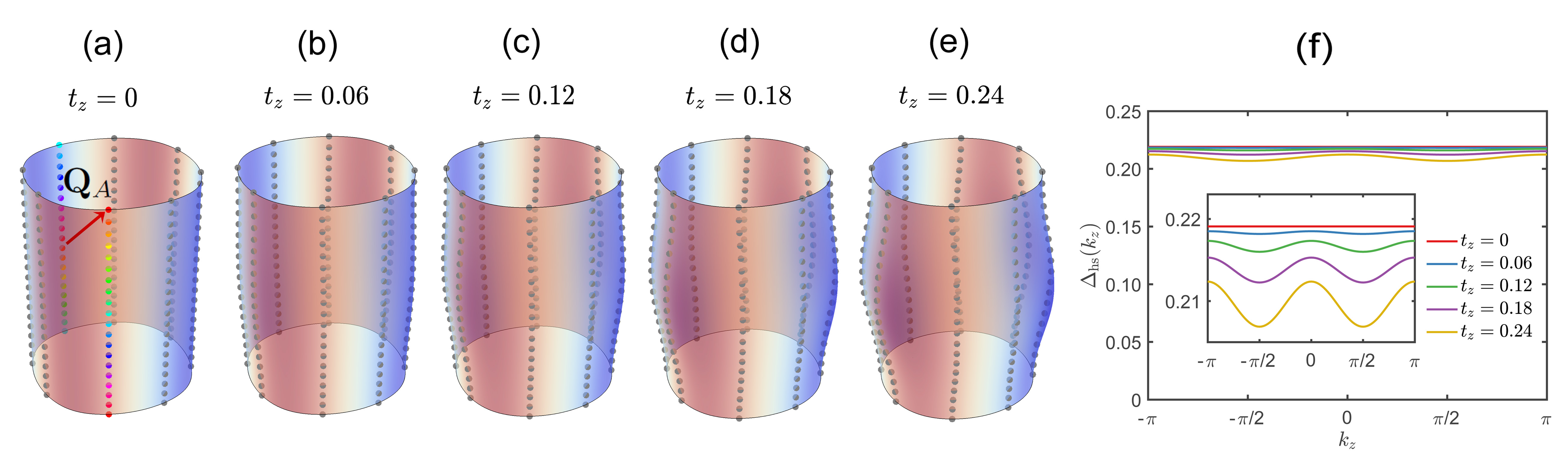}
\caption{Fermi surfaces and the hot lines associated with the wave-vector $\mbf{Q}_A=(\pi,\pi,\pi)$. The dispersion and parameters used here are identical to those in Fig.~\ref{fig:fs}. In panel (a), the hot spots connected by $\mbf{Q}_A$ are highlighted in the same color. Importantly, the hot lines are almost unaffected by increasing $t_z$, in contrast to the situation shown in Fig.~\ref{fig:fs} for the hot lines associated with $\mbf{Q}_M=(\pi,\pi,0)$. (f) Variation of the excitation gap along the hot lines associated with $\mbf{Q}_A$ for different values of $t_z$. The gap shows almost no variation as a function of $k_z$, in contrast to the case with $\mbf{Q}_M$ hot lines (Fig.~\ref{fig:hl1}). The inset zooms in a region to highlight the minor dependence on $k_z$.}
\label{fig:fs2}
\end{figure*}

Pairs of hot lines with opposite signs of the gap function emerge not only for $\mbf{Q}_M = (\pi,\pi,0)$, but for any momentum of the form $(\pi,\pi,Q_z)$. To investigate whether these momenta support a resonance mode, we consider now the dynamic susceptibility at the momentum corresponding to the magnetic wave-vector $\mbf{Q}_A = (\pi,\pi,\pi)$. The magnitude of the peak of the static bare susceptibility, $\text{Re}\chi_0(\mbf{q},\omega=0)$, at $\mbf{Q}_A$ is nearly identical to the peak at $\mbf{Q}_M$ (see Fig.~\ref{fig:chi0w0}). While for $\mbf{q}=\mbf{Q}_M$ the peak is slightly suppressed for increasing values of $t_z$, for $\mbf{q}=\mbf{Q}_A$, $t_z$ has no impact on the peak magnitude. 

In Fig.~\ref{fig:fs2}(a)--(e), we show the hot lines connected by the wave-vector $\mbf{Q}_A=(\pi,\pi,\pi)$. 
Of course, at $t_z=0$, the shape of the hot lines is unaffected by the addition of an out-of-plane component of the wave-vector. However, for larger values of $t_z$, shown in Figs.~\ref{fig:fs2}(b)--(e), the difference between the hot lines associated with $\mbf{Q}_M$ and $\mbf{Q}_A$ is prominent. In particular, in Fig.~\ref{fig:fs2}, the hot lines show almost no variation with $t_z$, in stark contrast to the case studied in Sec.~\ref{sec:qz0}, where the topology of the hot lines was affected by $t_z$, see Fig.~\ref{fig:fs}.

\begin{figure}
\includegraphics[width=0.9\columnwidth]{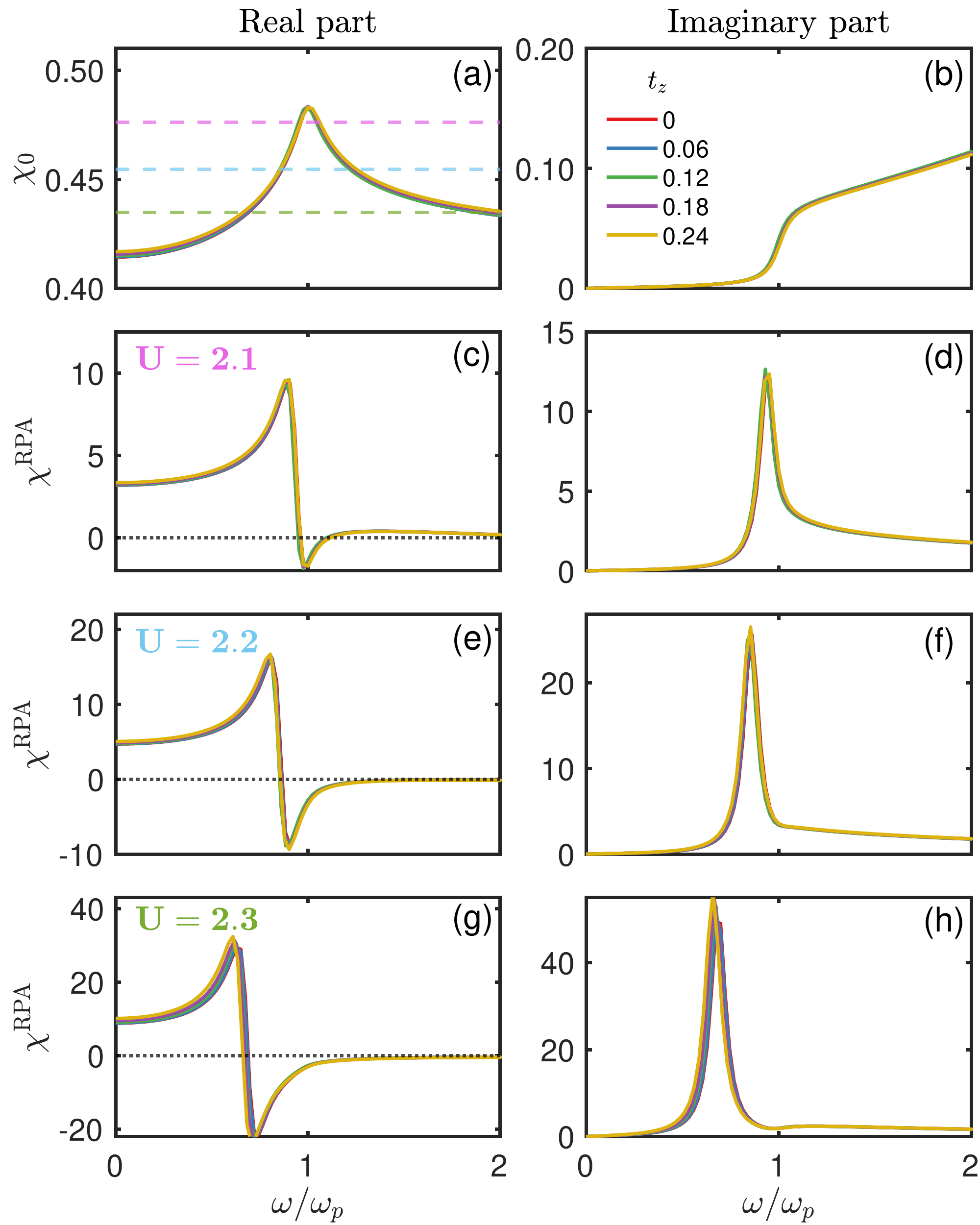}
\caption{Real (left column) and imaginary (right column) parts of the spin susceptibility at $\mbf{Q}_A=(\pi,\pi,\pi)$ for different values of $U$ and $t_z$. (a) and (b) show the real and imaginary parts of the bare susceptibility, respectively. As in Fig.~\ref{fig:chi0rpa}(a), the horizontal lines in (a) correspond to $1/U$ for progressively larger values of $U$. As in the $t_z=0$ case [panel (b)], the particle-hole continuum does not extend to zero frequency even for large values of $t_z$, in contrast to the $\mbf{Q}_M$ case shown in Fig.~\ref{fig:chi0rpa}. Hence, a prominent spin resonance mode appears for all values of $t_z$.}
\label{fig:chi0rpa2}
\end{figure}

As a result, the excitation gap along the hot lines shows almost no variation with $k_z$, as shown in Fig.~\ref{fig:fs2}(f). Therefore, exciting a collective mode with momentum $\mbf{Q}_A$ always incurs a finite energy cost, in contrast to collective modes with momentum $\mbf{Q}_M$, which can be excited with an  infinitesimally small energy above a certain value of $t_z$. Thus, in line with the reasoning in Sec.~\ref{sec:qz0}, for $\mbf{q}=\mbf{Q}_A$ we expect prominent spin resonance modes to appear for all values of $t_z$ between $t_z=0$ and $t_z=0.24$. This expectation is confirmed in Fig.~\ref{fig:chi0rpa2}, which shows the real and imaginary parts of the dynamical RPA susceptibility for $\mbf{q}=\mbf{Q}_A$.
These figures show clear resonance peaks whose shapes are unaffected by $t_z$. In contrast to the $\mbf{q}=\mbf{Q}_M$ case, the curves depicting $\text{Im}\chi_0(\mbf{Q}_A,\omega)$ in Fig.~\ref{fig:chi0rpa2}(b) are almost identical. This is related to the fact that, because the hot lines are essentially unchanged upon increasing $t_z$, the minimum and maximum excitation gaps along the hot lines, $\Delta_{\rm exc}$ and $\Delta_{\rm max}$, are very close to each other, such that $\omega_U$ is generally smaller than $\Delta_{\rm exc}$. This ensures that the resonance mode does not decay into the particle-hole continuum and remains sharp.

The gap, $\Delta_{\mbf{k}}$ of Eq.~\eqref{eq:gap_form}, can in principle depend on $k_z$ as well, in which case $\Delta_{\mbf{k}} \rightarrow (1+ \alpha \cos k_z a)\Delta_{\mbf{k}}$ for some value of $\alpha$. If $\alpha >1$, accidental line nodes are possible, which would inevitably intersect the hot lines and thus suppress the resonance mode. However, the $k_z$ dependence of the gap has been found to be weak for the parameters studied here~\cite{Roising2018weak}.

\subsection{Dispersion of the spin resonance mode along the $M-A$ line\label{sec:0topi}}

As we found in the previous sections, the absence or presence of a spin resonance mode is related to the topology of the hot lines. For a fixed momentum $\mbf{Q}_M$, there is a topological transition from open to closed hot lines as a function of increasing $t_z$. On the other other hand, for a fixed momentum $\mbf{Q}_A$, the hot lines remain open for all $t_z$ values studied. Therefore, for a fixed $t_z$ value, there can also be a topological transition from open to closed hot lines as a function of the out-of-plane momentum component $Q_z$ along the path from $M=(\pi,\pi,0)$ to $A=(\pi,\pi,\pi)$, which is parameterized by $\mbf{Q} = (\pi, \pi, Q_z)$ (see the Brillouin zone in the inset of Fig. \ref{fig:chi0w0}). This should have direct consequences for the dispersion of the spin resonance mode.

\begin{figure}
\includegraphics[width=\columnwidth]{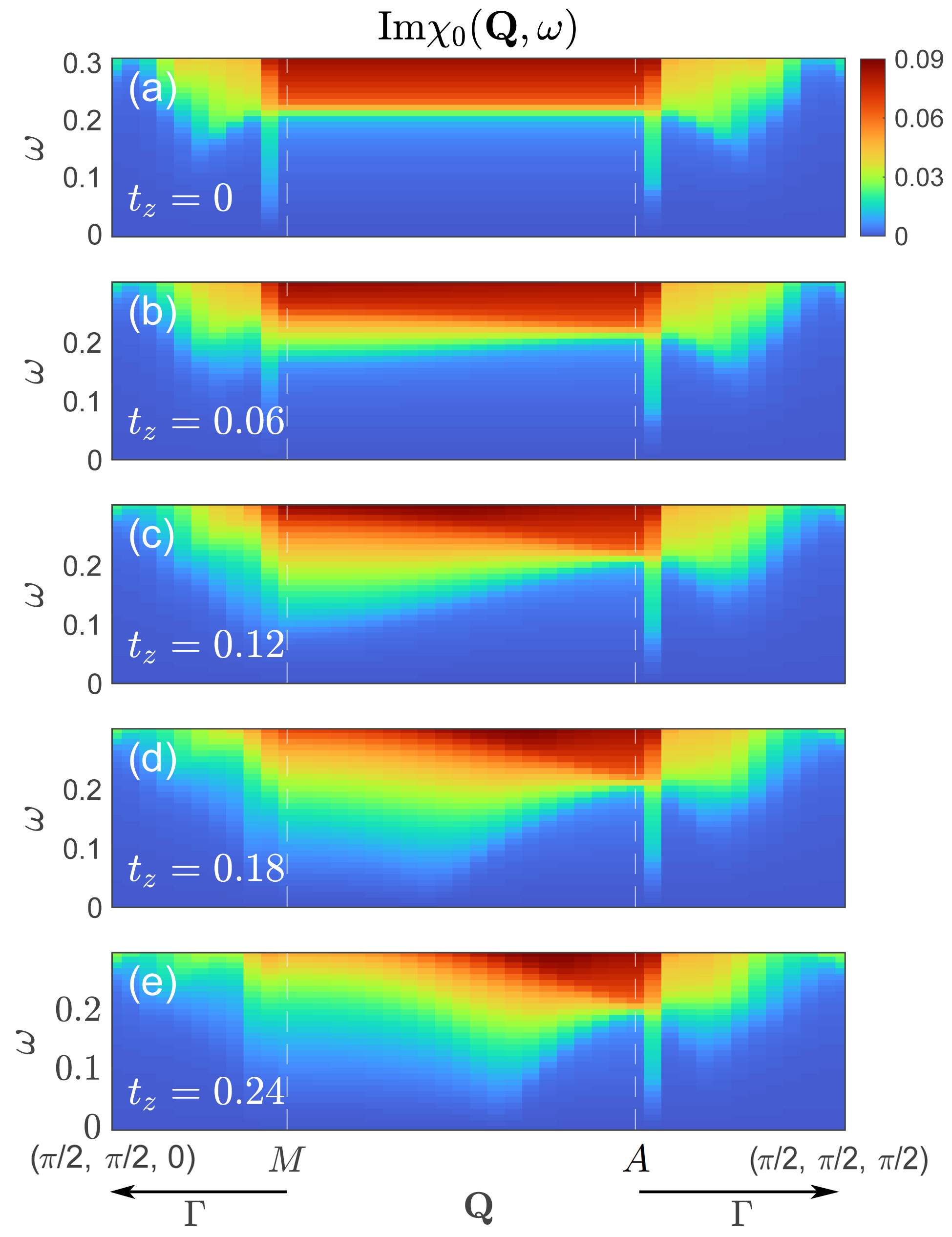}
\caption{Imaginary part of the bare susceptibility, $\text{Im}\chi_0(\mbf{q},\omega)$, as a function of $\omega$ for a high-symmetry path in the three-dimensional Brillouin zone focused on the $M-A$-direction. Here, a slightly smaller uniform grid with $1400^3$ points in momentum space was used. As $t_z$ is increased from (a) $t_z=0$ to (e) $t_z=0.24$, the particle-hole continuum extends to low energies over a progressively larger region near $M$, while it remains unaffected near $A$.}
\label{fig:chi0barecolorp}
\end{figure}
\begin{figure}
\includegraphics[width=\columnwidth]{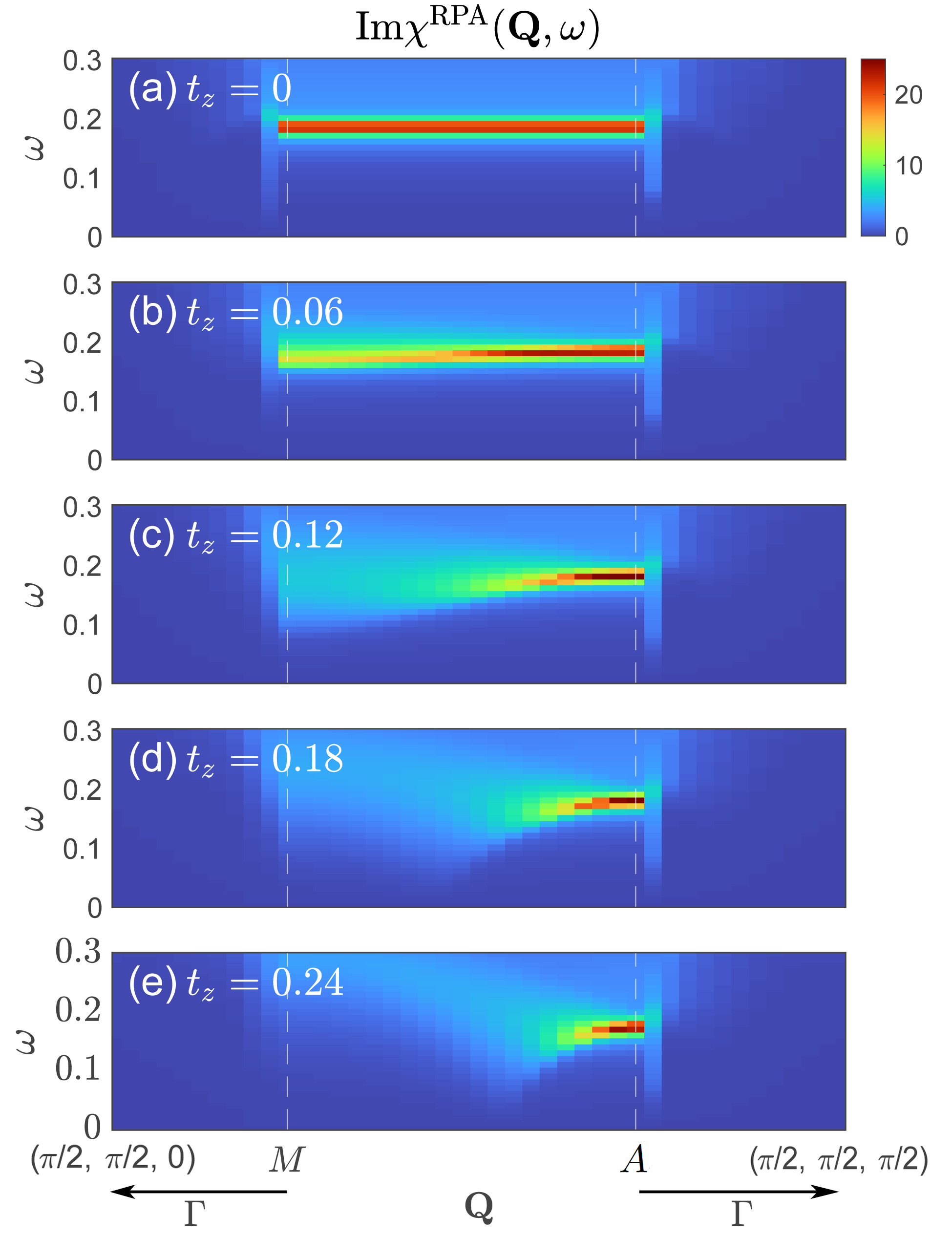}
\caption{Imaginary part of the RPA susceptibility, $\text{Im}\chi^{\rm RPA}(\mbf{q},\omega)$, as a function of $\omega$ for a high-symmetry path in the three-dimensional Brillouin zone focused on the $M-A$-direction for $U=2.2$. The corresponding bare susceptibility is shown in Fig.~\ref{fig:chi0barecolorp}. With the increase of $t_z$ from (a) $t_z=0$ to (e) $t_z=0.24$, the resonance mode acquires significant dependence on the out-of-plane momentum. In (a), the resonance mode exists over the entire path between $M$ and $A$, while in (e), it is restricted to a region surrounding $A$.}
\label{fig:colorp1}
\end{figure}
To investigate this problem, in Fig.~\ref{fig:chi0barecolorp} we show density plots of the imaginary part of the bare susceptibility, $\text{Im}\chi_0(\mbf{q},\omega)$, as function of $\omega$ and along a particular path of the Brillouin zone. Focusing on the path between $M$ and $A$, the evolution of the particle-hole continuum is clear. For $t_z=0$ [Fig.~\ref{fig:chi0barecolorp}(a)] there is a hard gap in the spectrum and no dependence on the out-of-plane momentum. However, as $t_z$ is increased, the particle-hole continuum moves towards smaller frequencies, with the suppression being the most prominent near $M$. This is a direct consequence of the fact that the minimum excitation gap $\Delta_{\rm exc}$ decreases significantly for the hot lines associated with small $Q_z$ values. As $t_z=0.18$ is reached and the particle-hole continuum near $M$ extends to zero frequency [Fig.~\ref{fig:chi0barecolorp}(d)], the out-of-plane momentum dependence of $\text{Im}\chi_0(\mbf{q},\omega)$ also becomes significant. Indeed, for $t_z=0.24$ [Fig.~\ref{fig:chi0barecolorp}(e)], the spectrum has a gap only in a narrow region near $A$, whose associated hot lines retain a non-zero $\Delta_{\rm exc}$ value.

As a result, we expect that the spin resonance mode will be overdamped along most of the $M-A$ line and will only persist near the $A$ point for moderate values of $t_z$. This is demonstrated in Fig.~\ref{fig:colorp1}: As anticipated, the resonance mode is sharp and non-dispersive along the $M-A$ line when $t_z=0$ [Fig.~\ref{fig:colorp1}(a)]. Upon increasing $t_z$, a $Q_z$ dispersion starts to develop. Eventually, for $t_z=0.24$ [Fig.~\ref{fig:colorp1}(e)], the resonance mode displays a clear dispersion as a function of $Q_z$, decaying into the particle-hole continuum at a momentum value close to $A$. Clearly, a sharp resonance is restricted to a narrow region near $A$. Finally, we remark on the seemingly discontinuous features appearing near $M$ and $A$ in Figs.~\ref{fig:chi0barecolorp} and \ref{fig:colorp1}. Finer momentum-grid calculations confirm that these features arise as a consequence of the rapid downward dispersion of the spin resonance, which is a well-known feature within the spin-exciton scenario~\cite{Eremin2005}. As $t_z$ is increased, this feature disappears near $M$, in agreement with the fact that no spin resonance is found there for larger values of $t_z$.

\section{Discussion and conclusions\label{sec:Conclusions}}

In this paper we systematically studied the impact of dimensionality on the spin resonance mode of unconventional superconductors. Starting from the 2D case, in Sec.~\ref{sec:2d} we reviewed the well-known conditions for the appearance of a spin-excitonic peak in the dynamical spin susceptibility of a $d_{x^2-y^2}$-wave superconductor \cite{Eschrig2006Effect,Abanov2003,scalapino2012}. As long as the hot spots of the Fermi surface are not fine-tuned to coincide with the nodes of the gap, the imaginary part of the bare susceptibility exhibits a hard excitation gap at zero temperature, as seen in Fig.~\ref{fig:2danalytic}(b). This fact, combined with the logarithmic divergence in the real part of the bare susceptibility, conspires to produce a well-defined spin resonance mode even for weak repulsive interactions.

A different situation arises in anisotropic three-dimensional $d$-wave superconductors, which we studied in detail in Sec.~\ref{sec:3d_cases}. A previous study by Chubukov and Gor'kov demonstrated that, for an isotropic system with a spherical Fermi surface, the logarithmic divergence present in 2D is replaced by a cusp, causing the spin resonance mode to disappear \cite{chubukov2008spin}. Here, we considered a model of layers coupled by a nearest-neighbor hopping term $t_z$. By restricting $t_z$ to be a fraction of the nearest-neighbor hopping $t$, we focused on layered systems with cylindrical-like but corrugated Fermi surfaces.
Upon systematically increasing $t_z$, and thus the corrugation of the Fermi surface, we showed that the hot lines associated with the in-plane magnetic wave-vector $\mbf{Q}_M=(\pi,\pi,0)$ undergo a topological transition from open to closed -- despite the fact that the Fermi surface remains open. When the hot lines form a closed loop, they cross the gap nodes at a specific $k_z$-plane  [see e.g. Fig.~\ref{fig:fs}(d)]. As a consequence, the behavior of $\text{Im}\chi_0(\mbf{Q}_M,\omega)$, as illustrated in Fig.~\ref{fig:wp}, is qualitatively different from the two-dimensional case, since it displays no hard gap. This causes the resonance mode to decay into the particle-hole continuum, thus becoming overdamped. In this case, the spin resonance mode at $\mbf{Q}_M$ is replaced by a broad incoherent peak instead. Moreover, for $t_z$ values near but below the threshold value for the hot-line topological transition, weak interactions are not enough to ensure the emergence of a spin-resonance mode, since the excitation gap can become very small.

The situation is markedly different for the magnetic spectrum at $\mbf{Q}_A=(\pi,\pi,\pi)$,  where a prominent spin resonance mode appears for all values of $t_z$ studied, as shown in Fig.~\ref{fig:chi0rpa2}. This is because, regardless of the corrugation of the Fermi surface, collective excitations with momentum $\mbf{Q}_A$ are subjected to a nearly $t_z$-independent gap, as shown in Fig.~\ref{fig:fs2}(f), which protects the spin resonance mode from damping by the particle-hole continuum. This is a direct consequence of the fact that the hot lines associated with $\mbf{Q}_A$ remain open for all $t_z$ values, as shown in Fig. \ref{fig:fs2}. The distinct behaviors of the resonance mode at $\mbf{Q}_M$ and $\mbf{Q}_A$ for moderate values of $t_z$ are manifested as a dispersive resonance mode along the $M-A$ path of the Brillouin zone, which is displayed in Fig.~\ref{fig:colorp1}.

While our calculations were done for a simple dispersion relation and a simple form of the $d_{x^2-y^2}$-wave superconducting gap function, our results are more general, as they ultimately rely on the topology of the hot lines and on the $d_{x^2-y^2}$ symmetry of the gap function. Whenever the hot lines associated with a magnetic wave-vector cross a nodal line, the spin-resonance mode should be strongly suppressed at the corresponding momentum. This is expected to happen across the topological transition of the hot lines from open to closed that is enabled by a reduced out-of-plane anisotropy. This is because neighboring hot lines for a $d_{x^2-y^2}$-wave state have different signs of the gap function, such that, when they close to form a loop, they will inevitably cross a nodal line. Note that the symmetry of the gap function is important for these conclusions to hold. For instance, if the gap had a $d_{xy}$-wave symmetry, the hot lines connected by the antiferromagnetic wave-vectors considered here would not have different gap signs.

The main implication of our results is that, even for layered $d_{x^2-y^2}$-wave superconductors displaying cylindrical-like Fermi surfaces in the normal state, a moderate out-of-plane anisotropy in the electronic dispersion (of about $20 \%$ in our model) can be enough to completely suppress the spin resonance mode for a wide range of momenta that are typically associated with antiferromagnetic fluctuations. This implies that a spin resonance mode can be absent in spite of a sign change in the superconducting order parameter. Thus, our results have clear implications for the interpretation of inelastic neutron scattering experiments in superconductors whose out-of-plane electronic dispersions are not negligible, which is the case for certain heavy-fermion compounds and possibly for the newly discovered nickelate NdNiO$_2$. Our results also point to the importance of probing the magnetic spectrum along the $(\pi,\pi,Q_z)$ direction to firmly establish the existence of a resonance mode.

\begin{acknowledgments}
We acknowledge useful discussions with A. V. Chubukov, A. Kreisel, and H. S. R{\o}ising. This work was supported by the U. S. Department of Energy through the University of Minnesota Center for Quantum Materials, under Grant No. DE-SC-0016371. The authors also acknowledge the Minnesota Supercomputing Institute at the University of Minnesota, where the numerical calculations were performed.  MHC also acknowledges support from a Carlsbergfondet fellowship.
\end{acknowledgments}

\bibliography{spinres}

\medskip{}

\end{document}